\documentclass[11pt]{article}

\usepackage[a4paper,margin=1in]{geometry}

\usepackage{graphicx}
\usepackage{amsmath}
\usepackage{amssymb}
\usepackage{authblk}
\usepackage{hyperref}
\hypersetup{
  colorlinks=true,
  linkcolor=blue,
  citecolor=blue,
  urlcolor=blue
}

\title{Hybrid-Contact Planar HPGe Process Vehicle Toward Ring-Contact Designs}

\author[1]{Kunming Dong}
\author[1]{Dongming Mei\thanks{Corresponding author: \texttt{dongming.mei@usd.edu}}}
\author[1]{Shasika Panamaldeniya}
\author[1]{Anupama Karki}
\author[1]{Patrick Burns}
\author[1]{Sanjay Bhattarai}

\affil[1]{Department of Physics, University of South Dakota, Vermillion, SD 57069, USA}

\date{}

\begin{document}

\maketitle

\begin{abstract}
Rare-event searches—including dark matter, coherent elastic neutrino--nucleus scattering (CE$\nu$NS), and neutrinoless double-beta decay (0$\nu\beta\beta$)—require high-purity germanium (HPGe) detectors with ultralow noise, stable backgrounds, and electrode geometries that can scale to larger single-crystal masses. Ring-contact (ring-and-groove) designs address scalability by shaping the electric field to preserve low-capacitance readout, but their nonplanar topology motivates a lithium-contact process that is compatible with conformal deposition and robust high-voltage operation. As a process demonstration toward future ring-contact prototypes, we fabricate and characterize a hybrid-contact planar HPGe device, KL01. Here, ``hybrid'' denotes an $n^{+}$ contact formed by an in-house lithium-suspension paint followed by controlled thermal diffusion, combined with an AJA-developed a-Ge/Al $p^{+}$ contact and a-Ge sidewall passivation. At 77~K the device exhibits pA-scale leakage current under kV bias, a depletion plateau near $V_{\mathrm{dep}}\approx 1300$~V, and energy resolutions of 1.57~keV FWHM at 59.5~keV and 2.57~keV FWHM at 662~keV. These results validate the compatibility of the paint-and-diffuse lithium process with thin-film a-Ge/Al contacts and establish a practical fabrication workflow to be extended to ring-and-groove electrodes for next-generation rare-event HPGe modules.


\end{abstract}

\section{Introduction}
High-purity germanium (HPGe) detectors are central to rare-event searches and precision radiation spectroscopy because they combine excellent energy resolution with very low electronic noise~\cite{Wei2020_EPJC_CBH,Panth2020_EPJC_Cryo}. For neutrinoless double-beta decay ($0\nu\beta\beta$) and related low-background programs, point-contact-style electrodes have been particularly impactful. Broad Energy Germanium (BEGe) detectors in GERmanium Detector Array (GERDA) and p-type point-contact (PPC) detectors in the Majorana Demonstrator demonstrated strong pulse-shape discrimination (PSD), enabling efficient separation of single-site energy depositions (signal-like) from multi-site interactions (background-like) and thereby improving background rejection in $^{76}$Ge searches~\cite{Agostini2013_GERDA_PSD,Alvis2019_MJD_MSE,Agostini2018_GERDA_PhaseII_PRL,Arnquist2023_MJD_Final}. In practice, however, these BEGe/PPC devices are typically at the $\sim$1~kg scale; scaling a pure point-contact geometry to substantially larger masses becomes increasingly challenging because full depletion and favorable field/weighting-potential conditions over the entire crystal can demand impractically high bias or can leave partially depleted regions that degrade charge collection and PSD performance.

The inverted coaxial point-contact (ICPC) geometry preserves the PSD advantages of point-contact readout while enabling larger crystals through field shaping and a reduced surface-to-volume ratio~\cite{Cooper2012_NIMA_HPGeTrackingImaging,Agostini2021_EPJC_ICPC,Comellato2021_EPJC_Collective}. Consequently, the Large Enriched Germanium Experiment for Neutrinoless $\beta\beta$ Decay (LEGEND) deploys ICPC detectors in LEGEND-200, and LEGEND-1000 plans to instrument on the order of $\sim$400 ICPC units with typical individual masses of 2--3~kg to reach an active mass of 1000~kg~\cite{Agostini2021_EPJC_ICPC,Abgrall2021_LEGEND1000}. Increasing the mass per detector remains attractive because it can reduce channel count and cabling, simplify mechanical integration, and lower the total detector surface area that can host backgrounds.

A promising route to even larger single-crystal masses is the germanium ring-contact geometry proposed by Dr.~David Radford, which uses a ring-and-groove electrode topology to shape the electric field while maintaining low capacitance and point-contact-like signal formation~\cite{Radford2024_RingContactTalk,Leone2022_Thesis_RingContact,Radford2018_PIRE_RingContact}. Field simulations indicate that ring-contact designs can extend the scalable mass to multi-kg crystals, potentially approaching $\sim$6~kg at bias voltages of a few kV (comparable to ICPC operation), offering a compelling path to further reduce detector count and complexity in LEGEND-1000 and beyond. The main fabrication challenge is that the recessed ring and groove topology complicates uniform one-step lithium evaporation and subsequent diffusion on large crystals: achieving continuous, conformal coverage on vertical sidewalls and around grooves is nontrivial.

Accordingly, further improvements in scalability and background performance are expected from refinements within the lithium-diffused n$^+$ framework, including more uniform diffusion profiles, stable passivation around sidewalls and grooves, and contact metallization that maintains low capacitance in large crystals~\cite{Ma2017_ARI_InactiveLayer,Dai2022_arXiv_LiInactive}. Amorphous semiconductor contacts are valuable for small detectors and development studies~\cite{Wei2019_JINST_aGeDetectors,Bhattarai2020_EPJC_aGeConduction,Wei2022_EPJC_PlanarPPC}. Although vacuum evaporation is generally preferred for lithium deposition in hybrid planar devices~\cite{Hull2022_PHDS_RCD}, in this work we adopt an in-house lithium suspension paint with a controlled diffusion schedule as a process pathfinder for ring-contact fabrication~\cite{Radford2024_RingContactTalk,Leone2022_Thesis_RingContact,Radford2018_PIRE_RingContact}. Paint-on deposition followed by controlled diffusion can provide conformal lithium coverage on nonplanar features, including vertical sidewalls and recessed grooves. Demonstrating low-leakage performance on a compact planar test detector therefore reduces technical risk and establishes a practical route to larger-mass ring-contact detectors.

In recent years, amorphous semiconductor contacts such as amorphous germanium and amorphous silicon have become a reliable option for thin, passivating electrodes on HPGe detectors~\cite{Panth2020_EPJC_Cryo,Wei2019_JINST_aGeDetectors,Amman2020_SegmentedAS}. When properly prepared, amorphous germanium contacts provide strong and approximately symmetric blocking barriers for both holes and electrons, enabling low-leakage planar devices and allowing reversal of the bias polarity and depletion direction without degrading performance~\cite{Wei2020_EPJC_CBH,Bhattarai2020_EPJC_aGeConduction,Luke1992_aGeBipolar}. At the University of South Dakota (USD), we purify commercial Ge ingots to detector-grade quality via multi-pass zone refining~\cite{Yang2014_CRT_ZoneRefine,Yang2015_JPCS_ZoneRefine}, grow high-purity single crystals by the Czochralski method and characterize them for detector applications~\cite{Wang2015_JPCS_GeGrowth,Raut2020_JINST_CrystalChar,Bhattarai2024_Crystals_Growth}, and fabricate small planar HPGe test detectors with sputtered amorphous-germanium contacts from these USD-grown crystals~\cite{Wei2019_JINST_aGeDetectors,Wei2022_EPJC_PlanarPPC,Mei2020_JPG_ChargeTrapping}. Scaling to larger detectors required new equipment and systematic process optimization. In 2023, we installed an AJA magnetron sputtering system to support larger samples and tighter control of film properties. After roughly two years of development, we identified a stable recipe that yields good adhesion, reproducible resistivity, and low interface charge. Recently, we demonstrated a bipolar planar device produced on this system with a base of approximately 20~$\times$~21~mm, a top of approximately 13~$\times$~14~mm, and a thickness of 10~mm. In this work, a ``bipolar'' detector denotes a planar p-type HPGe device with thin amorphous-germanium blocking contacts on both electrodes (i.e., without a thick Li-diffused dead layer). That device reached full depletion at about 1400~V, showed an impurity concentration of approximately $2.6\times10^{10}$~cm$^{-3}$, and exhibited leakage currents on the order of 10~pA at 77~K. This result provided the technical basis and confidence to proceed to the present device.

In this work, a ``hybrid-contact'' detector denotes a mixed contact architecture that combines a conventional Li-diffused n$^+$ electrode with a thin amorphous blocking electrode (a-Ge/Al) and a-Ge sidewall passivation, thereby distinguishing it from (i) junction-style contacts (Li n$^+$ paired with an implanted/diffused p$^+$ electrode) and (ii) fully amorphous-contact schemes (a-Ge/a-Si blocking contacts on both electrodes). The new element here is the controlled integration of our in-house lithium paint/diffusion protocol with an AJA-validated a-Ge/Al process on a compact planar test vehicle; the accepted trade-off is retaining a thick dead layer on the Li side while keeping the opposite electrode thin and conformal for field shaping.

In this context, we present the hybrid planar HPGe detector KL01 as the next step in our development path. Our aim is to qualify the lithium paint and diffusion protocol alongside the AJA-based a-Ge/Al recipe before moving to the ring-contact geometry proposed by Dr.~David Radford~\cite{Leone2022_Thesis_RingContact,Radford2018_PIRE_RingContact,Hull2022_PHDS_RCD}. Demonstrating stable depletion, low leakage, and competitive energy resolution in KL01 establishes the fabrication basis required for uniform contacts on ring and groove features. The sections that follow describe the materials and methods, the contact architecture, and the electrical performance of KL01, and they outline the remaining steps toward larger-mass ring-contact detectors for next-generation experiments.

\section{Process-Validation Objectives and Path to Ring-Contact Scale-Up}
The goal of this study is to fabricate and qualify a full-function hybrid-contact planar HPGe detector (KL01) as a physics-driven process-validation milestone for the University of South Dakota (USD) detector pipeline. Hybrid-contact architectures intentionally combine a thick, lithium-diffused n$^+$ electrode (robust, mature, high-voltage tolerant) with a thin, sputtered amorphous-semiconductor blocking contact on the opposite face (low capacitance and stable surface passivation), enabling a practical balance between operability and field shaping on p-type crystals~\cite{Wei2020_EPJC_CBH,Wei2019_JINST_aGeDetectors,Luke1992_aGeBipolar,Knoll2010_RDM,Canberra2017_GeManual}. From a device-physics standpoint, the validation must demonstrate (i) controlled donor formation and stability of the Li-diffused contact, (ii) reliable carrier-injection blocking and low surface leakage from the amorphous-contact and passivation scheme, and (iii) predictable bulk depletion behavior consistent with the net impurity concentration and detector electrostatics.

On the n$^+$ side, lithium diffusion into Ge is governed by temperature-dependent diffusivity and solid solubility, and is complicated by precipitation and time-dependent redistribution that can modify the effective donor profile and the thickness of the Li-related inactive layer~\cite{Fuller1953_LiDiff_GeSi,Pell1957_LiSolub_Ge,Morin1957_LiPrecip_Ge,Wenzl1971_LiPrecipitates_Ge,Crank1975_Diffusion,Shewmon1989_Diffusion}. These phenomena are directly relevant to p-type point-contact style detectors because the near-contact region can exhibit incomplete charge collection and pulse-shape/energy-degradation effects; therefore, controlling the Li profile and maintaining uniformity are essential for preserving both spectroscopy and background-rejection performance~\cite{Ma2017_ARI_InactiveLayer,Dai2022_arXiv_LiInactive,Aguayo2013_NIMA_LiNplus,Andreotti2014_ARI_DeadLayer}. In this work, the lithium-suspension ``paint'' route is treated as a conformal-deposition strategy intended to improve coverage on nonplanar features; its qualification on a planar vehicle reduces fabrication risk prior to implementing recessed grooves and rings.

On the p-side, amorphous-semiconductor contacts (a-Ge/a-Si) are widely used in HPGe R\&D because they can provide thin, conformal, passivating electrodes with approximately symmetric blocking for electron and hole injection when properly prepared~\cite{Wei2019_JINST_aGeDetectors,Bhattarai2020_EPJC_aGeConduction,Amman2020_SegmentedAS,Luke1992_aGeBipolar}. At cryogenic temperature, leakage current is often dominated by carrier injection at imperfect contacts and by surface leakage along sidewalls; thus validating the sputtered a-Ge/Al contact and a-Ge passivation is fundamentally a validation of barrier formation, interface stability, and surface conduction suppression~\cite{Wei2020_EPJC_CBH,Panth2020_EPJC_Cryo,Bhattarai2020_EPJC_aGeConduction}. These issues become more stringent as detector size increases because higher bias is required to deplete larger volumes and because surfaces and field nonuniformities play a larger role in leakage and noise.

Accordingly, the specific objectives of KL01 are:

\begin{itemize}
\item \textbf{Lithium n$^+$ formation and uniformity.}
Establish a reproducible Li-diffused n$^+$ backside contact on a USD-grown crystal using the lithium-suspension paint and a controlled thermal diffusion schedule. The physics requirement is that the resulting donor profile produces a stable n$^+$/p transition region and an inactive-layer behavior consistent with Li diffusion/solubility constraints and known precipitation phenomena~\cite{Fuller1953_LiDiff_GeSi,Pell1957_LiSolub_Ge,Morin1957_LiPrecip_Ge,Wenzl1971_LiPrecipitates_Ge,Crank1975_Diffusion,Shewmon1989_Diffusion}. Device-level implications include stable high-voltage operation and minimized charge-collection degradation from near-contact events ~\cite{Ma2017_ARI_InactiveLayer,Dai2022_arXiv_LiInactive,Aguayo2013_NIMA_LiNplus}.

\item \textbf{Blocking-contact quality and stable injection suppression.}
Produce a uniform a-Ge/Al signal contact on the opposite face by sputtering using the AJA-validated recipe previously demonstrated on USD planar devices, and confirm contact quality through I--V/C--V behavior. The key physics metric is low carrier-injection current at 77~K (effective blocking barriers) and stable behavior under increasing reverse bias, consistent with established amorphous-Ge contact performance and conduction models~\cite{Wei2020_EPJC_CBH,Wei2019_JINST_aGeDetectors,Bhattarai2020_EPJC_aGeConduction,Wei2022_EPJC_PlanarPPC,Luke1992_aGeBipolar}.

\item \textbf{Surface passivation and leakage suppression.}
Implement high-quality a-Ge sidewall passivation to suppress surface leakage paths and enable stable high-voltage operation. Since surface conduction and interface charge can dominate leakage and noise in planar geometries, the passivation objective is assessed by the absence of surface-driven leakage instabilities and by reproducible I--V behavior at cryogenic temperature~\cite{Panth2020_EPJC_Cryo,Bhattarai2020_EPJC_aGeConduction,Amman2020_SegmentedAS}.

\item \textbf{Bulk depletion physics and electrostatics validation.}
Demonstrate full depletion at 77~K at a stable operating bias, with depletion behavior consistent with the measured net impurity concentration and planar electrostatics. Practically, the depletion transition is verified via C--V trends and/or pulser-amplitude response and cross-checked against standard depletion-voltage methods used for HPGe characterization~\cite{Knoll2010_RDM,Canberra2017_GeManual,Mirion_HPGeLab,Abt2021_JINST_SSD}.

\item \textbf{Noise and spectroscopy performance.}
Demonstrate low leakage current over the operating range and obtain excellent energy resolution using standard $\gamma$-ray sources. The physics interpretation connects resolution to the combined effects of electronic noise (capacitance, leakage-current shot noise), charge collection, and contact stability~\cite{Wei2020_EPJC_CBH,Panth2020_EPJC_Cryo,Knoll2010_RDM}.

\item \textbf{Process controls and metrology for scale-up.}
Document process controls (cleaning, etching, contamination mitigation) and measurement protocols (I--V, C--V, spectroscopy) so that outcomes are repeatable and transferable to larger crystals and more complex electrode topologies. This includes identifying critical steps that influence contact barrier formation, surface leakage, and Li inactive-layer behavior~\cite{Ma2017_ARI_InactiveLayer,Dai2022_arXiv_LiInactive,Andreotti2014_ARI_DeadLayer}.
\end{itemize}

\noindent\textbf{Path to ring-contact detectors.}
The immediate motivation for KL01 is to de-risk the contact and passivation steps that are most challenging to implement on nonplanar electrode topologies. The ring-contact geometry proposed by Radford introduces recessed grooves and ring features that must be uniformly coated and subsequently diffused to form a continuous n$^+$ layer while preserving low capacitance and point-contact-like signal formation~\cite{Radford2024_RingContactTalk,Leone2022_Thesis_RingContact,Radford2018_PIRE_RingContact}. Because one-step lithium evaporation can struggle to deliver uniform coverage on vertical sidewalls and within grooves, the paint-and-diffuse method is pursued specifically for its conformality. Demonstrating that this approach yields stable depletion, low leakage, and competitive resolution on a planar process vehicle establishes the fabrication basis for scaling to ring-and-groove electrodes. Ultimately, successful translation to ring-contact designs would enable larger-mass, low-capacitance HPGe modules and reduce detector-count complexity for next-generation rare-event searches, complementing the ICPC strategy used in LEGEND-200 and envisioned for LEGEND-1000~\cite{Agostini2021_EPJC_ICPC,Abgrall2021_LEGEND1000}.

\section{Detector Design and Fabrication}
\label{sec:design_fab}
\begin{figure}[!htbp]
  \centering
  \includegraphics[width=0.6\linewidth]{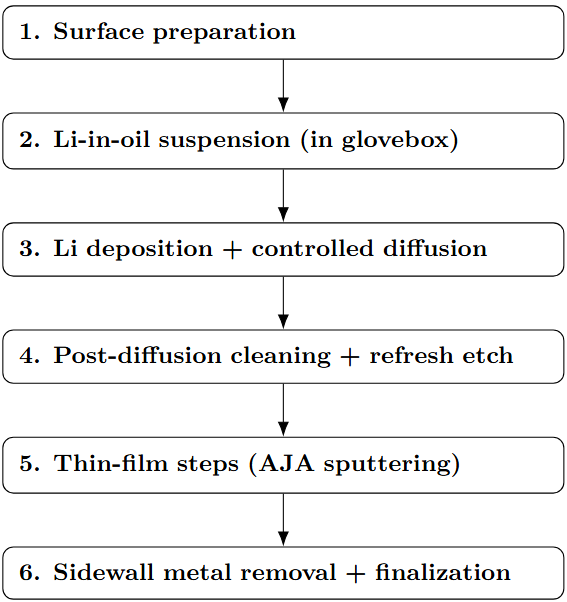}
  \caption{Simplified fabrication process-flow diagram for the KL01 hybrid-contact planar HPGe detector. The detailed procedures for each stage are described in Secs.~3.4--3.9.}
  \label{fig:kl01_process_flow}
\end{figure}

The KL01 detector was fabricated from a high-purity Ge single crystal grown at the University of South Dakota (USD) by the Czochralski (CZ) method, following the USD purification--growth--characterization pipeline reported in prior work~\cite{Yang2014_CRT_ZoneRefine,Yang2015_JPCS_ZoneRefine,Wang2015_JPCS_GeGrowth,Raut2020_JINST_CrystalChar,Bhattarai2024_Crystals_Growth}. For clarity, a simplified process-flow diagram of the KL01 fabrication is shown in Fig.~\ref{fig:kl01_process_flow}. This schematic is intended only as a high-level roadmap of the major fabrication stages; the complete, step-by-step procedures and experimental details corresponding to each stage are provided in Secs.~3.4--3.9.

The material is p-type with a net acceptor concentration on the order of $N_{\mathrm{eff}}\approx 3.14\times10^{10}$~cm$^{-3}$, consistent with detector-grade HPGe where full depletion can be achieved at kV-scale bias for cm-scale thicknesses~\cite{Knoll2010_RDM,Haller1981_PhysicsUPGe}. The crystal was machined into a planar geometry with approximate face dimensions of $10.34~\mathrm{mm}\times 11.28~\mathrm{mm}$ and a thickness of $d=9.06$~mm. All faces were mechanically lapped and polished, then chemically etched to remove the mechanically damaged layer and reduce surface contamination prior to contact formation~\cite{Wei2019_JINST_aGeDetectors,Amman2020_SegmentedAS}.

\subsection{Hybrid-contact architecture and electrical configuration}
\label{subsec:contact_architecture}
\begin{figure}[!htbp]
  \centering
  \includegraphics[width=0.8\columnwidth]{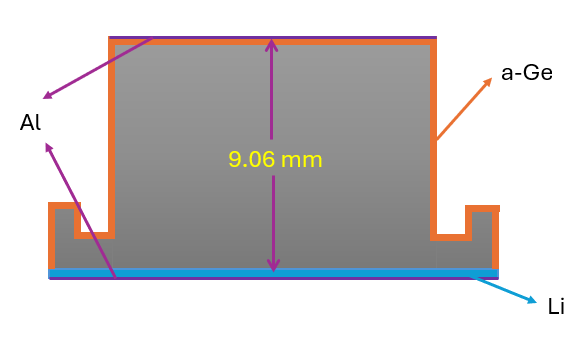}
  \caption{Cross-sectional schematic of the hybrid planar HPGe detector.}
  \label{fig:detector_cross_section}
\end{figure}

KL01 uses a \emph{hybrid-contact} scheme (Fig.~\ref{fig:detector_cross_section}): (i) a lithium-diffused n$^+$ backside electrode that forms a robust, high-voltage-tolerant contact but introduces a thick Li-related inactive layer, (ii) a thin sputtered amorphous-germanium layer capped with Al on the opposite face that functions as the p-side signal electrode, and (iii) a-Ge sidewall passivation without a wrap-around electrode. Hybrid architectures of this type are widely used in HPGe development because they combine the operational robustness of Li-diffused n$^+$ contacts with the conformality and surface-stability advantages of amorphous-semiconductor blocking/passivating layers~\cite{Knoll2010_RDM,Canberra2017_GeManual,Luke1992_aGeBipolar,Wei2019_JINST_aGeDetectors,Amman2020_SegmentedAS}. In particular, amorphous-Ge contacts can provide effective carrier-injection blocking at cryogenic temperature (reducing leakage current and shot-noise contributions) while remaining thin and conformal on planar and moderately nonplanar surfaces~\cite{Wei2020_EPJC_CBH,Panth2020_EPJC_Cryo,Bhattarai2020_EPJC_aGeConduction,Luke1992_aGeBipolar}.

In operation, the high voltage is applied to the n$^+$ backside electrode, while the a-Ge/Al top contact is held near ground and connected to a charge-sensitive preamplifier. This configuration is chosen to preserve low effective input capacitance and to maintain stable readout at 77~K, where electronic noise is strongly influenced by capacitance and leakage-current shot noise~\cite{Wei2020_EPJC_CBH,Knoll2010_RDM}. The sidewalls are passivated with a-Ge only; deliberate removal of sidewall metal prevents unintended wrap-around electrodes that can increase capacitance and introduce surface leakage paths under high field~\cite{Wei2019_JINST_aGeDetectors,Amman2020_SegmentedAS}.

\subsection{Mechanical handling features and field-shaping considerations}
\label{subsec:handling_features}
To protect active surfaces during wet processing and cryostat assembly, the planar blank was cut with (i) a $\sim$2~mm handling wing and (ii) a $\sim$2~mm perimeter groove. These features allow tools to contact only the wing and establish a ``no-touch'' margin defined by the groove. The groove also helps suppress accidental metal wrap-around and mitigates the formation of unintended conductive paths along edges during subsequent metallization and etching steps~\cite{Amman2020_SegmentedAS}. In KL01, both features are outside the primary drift region defined by the overlapping electrodes; thus their influence on the active capacitance and the near-depletion field uniformity is expected to be small. This expectation is verified by electrostatic simulation (Sec.~\ref{subsec:efield_sim}), which shows that field lines relevant to bulk charge transport are primarily confined to the overlap region.

\subsection{Electrostatic simulation and depletion-voltage estimate}
\label{subsec:efield_sim}
\begin{figure}[!htbp]
  \centering
  \includegraphics[width=0.8\columnwidth]{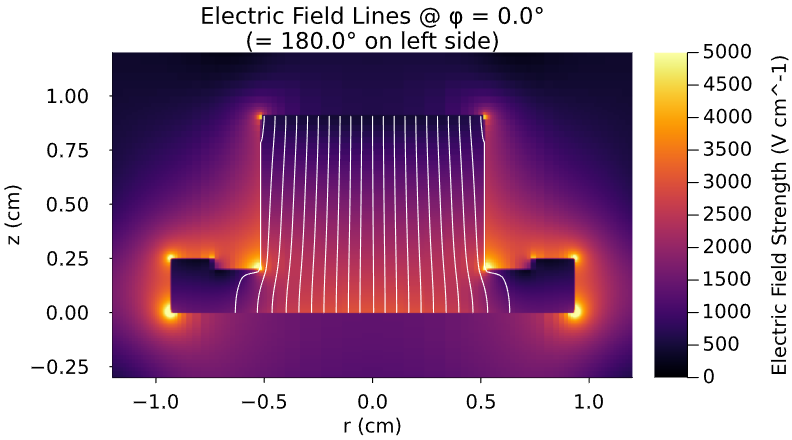}
  \caption{Simulated electric-field magnitude and field lines for the realized hybrid planar HPGe detector geometry at an applied bias of 1600~V, obtained with SolidStateDetectors.jl (SSD.jl)~\cite{Abt2021_JINST_SSD}.}
  \label{fig:detector_efield}
\end{figure}

Figure~\ref{fig:detector_efield} shows the simulated electric-field magnitude and field lines for the realized geometry using SolidStateDetectors.jl (SSD.jl)~\cite{Abt2021_JINST_SSD}. As a baseline analytic estimate, the depletion voltage of a planar detector with uniform net space-charge density can be approximated by~\cite{Knoll2010_RDM}
\begin{equation}
V_{\mathrm{dep}} \approx \frac{e\,N_{\mathrm{eff}}\,d^{2}}{2\,\varepsilon_{\mathrm{Ge}}}\,,
\label{eq:vdep_planar}
\end{equation}
where $e$ is the elementary charge, $N_{\mathrm{eff}}$ is the net acceptor concentration (for p-type), $d$ is the active thickness, and $\varepsilon_{\mathrm{Ge}}$ is the permittivity of Ge. Using $N_{\mathrm{eff}}\approx 3.14\times10^{10}$~cm$^{-3}$ and $d=9.06$~mm yields $V_{\mathrm{dep}}\approx 1.46$~kV for an ideal planar diode (neglecting dead-layer effects and non-ideal boundary conditions). In practice, the presence of a Li-related inactive layer and fringing fields near edges can shift the apparent depletion behavior, motivating the combined use of simulation and C--V (or pulser-proxy) measurements to identify the operational plateau~\cite{Knoll2010_RDM,Mirion_HPGeLab,Abt2021_JINST_SSD}.

\subsection{Surface preparation and contamination control prior to contact formation}
\label{subsec:surface_prep}
Fabrication was carried out in USD's semiconductor detector laboratory using procedures developed for amorphous contact HPGe detectors~\cite{Wei2019_JINST_aGeDetectors,Wei2022_EPJC_PlanarPPC}. After cutting and lapping, a heavy chemical etch with HNO$_3$:HF (4:1 by volume) was used to remove the damaged near-surface layer and reduce contamination introduced by mechanical processing~\cite{Wei2019_JINST_aGeDetectors,Amman2020_SegmentedAS}. Following etching, the detector was blow-dried with high-purity nitrogen and transferred into an argon glovebox through an antechamber cycle (pump and backfill). Low-oxygen conditions are essential because lithium and freshly etched Ge surfaces are chemically reactive; uncontrolled oxidation or contamination can degrade contact uniformity, increase leakage, and reduce the reproducibility of subsequent diffusion and sputtering steps~\cite{Wei2019_JINST_aGeDetectors,Amman2020_SegmentedAS,Yin1969_Thesis_LiPaint}. Inside the glovebox, the detector was placed on an isopropyl-alcohol (IPA)-cleaned ultrafine-grain graphite plate and positioned on the internal heater platform for lithium deposition and diffusion.

\subsection{Lithium-in-oil suspension preparation}
\label{subsec:li_suspension}
A lithium-in-oil suspension was required to enable conformal ``paint-on'' lithium deposition. Open formulations are not available from industry, and the only public description we identified is a 1969 thesis reporting a 30~wt\% lithium-in-oil mixture using 10--30~$\mu$m lithium powder~\cite{Yin1969_Thesis_LiPaint}. Because high-purity lithium powders and ready-made suspensions are no longer commercially accessible and vendor sourcing is not disclosed, we developed an in-house preparation method.

All preparation was performed inside an argon glovebox. High-purity mineral oil was placed in a ceramic crucible and lithium foil (20~$\mu$m thickness) was fractured directly into the oil using high-power ultrasonication. Sonication was performed with a 600~W ultrasonic homogenizer (20~kHz) equipped with a 16~mm titanium-alloy probe. Lithium foil was added in $\sim$0.1~g portions per cycle, ultrasonicated for 2~min, and allowed to cool for 1~h. This add--sonicate--cool sequence was repeated until the slurry became noticeably more viscous. Above $\sim$15~wt\% lithium, energy coupling degraded and splashing was observed, setting a practical upper bound for direct sonication in our geometry.

\begin{figure}[!htbp]
  \centering
  \includegraphics[width=0.8\columnwidth]{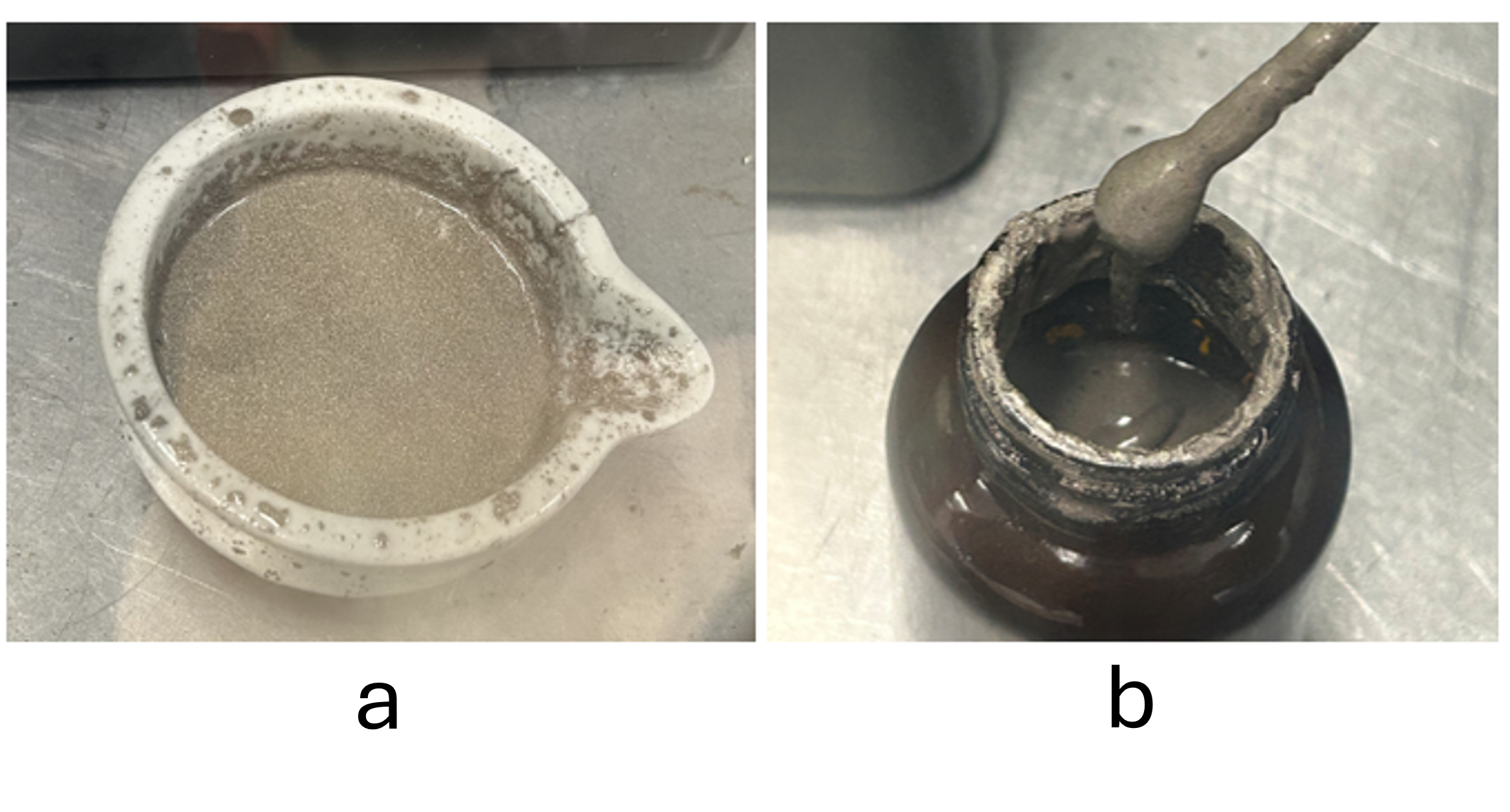}
  \caption{Lithium-in-oil preparation stages. Panel (a) shows the lithium-in-oil mixture in a ceramic crucible during ultrasonication, and panel (b) shows the concentrated lithium-in-oil suspension transferred to an amber glass bottle for storage prior to use.}
  \label{fig:li_paint_prep}
\end{figure}

To reach a target near 30~wt\%, excess mineral oil was removed by gravity separation. Because lithium is less dense than mineral oil, a lithium-rich phase accumulates at the top after settling. The top concentrated layer was skimmed with a ceramic spoon and transferred to an amber glass bottle (Fig.~\ref{fig:li_paint_prep}), yielding a measured lithium mass fraction of $\approx$28~wt\%. A coarse particle-size assessment by optical microscopy (50$\times$) showed that most particles were below $\sim$10~$\mu$m, with rare outliers below $\sim$40~$\mu$m, comparable to the particle-size range discussed in Ref.~\cite{Yin1969_Thesis_LiPaint}. The suspension was sealed and stored under argon; visual monitoring of metallic luster was used as a practical indicator of chemical stability (loss of luster or discoloration is consistent with degradation due to oxidation/nitridation)~\cite{Yin1969_Thesis_LiPaint}.

\subsection{Lithium deposition and controlled diffusion}
\label{subsec:li_diffusion}
Prior to use, the suspension was homogenized by stirring to re-disperse lithium in the oil. A small amount was loaded onto a fine brush and applied as a thin, uniform film on the designated Ge face. Film uniformity is critical: insufficient lithium can yield a poor diode, while excess lithium can produce surface pitting and nonuniform diffusion after thermal processing~\cite{Yin1969_Thesis_LiPaint}. In our experience, an appropriate loading produces a uniform brick-red appearance after diffusion, consistent with the qualitative indicators described in Ref.~\cite{Yin1969_Thesis_LiPaint}.

\begin{figure}
  \centering
  \includegraphics[width=0.8\columnwidth]{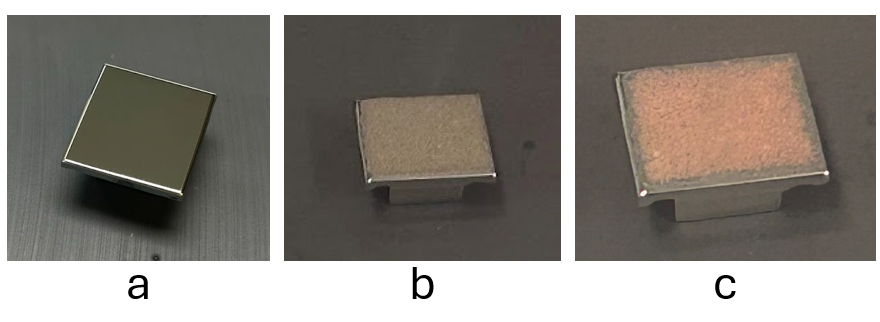}
  \caption{Evolution of the detector bottom surface during the lithium-paint diffusion process: (a) detector on a graphite plate inside the glovebox before lithium deposition; (b) detector after application of lithium paint; (c) detector after diffusion and cooling with uniform brick-red appearance.}
  \label{fig:LiPaintStages}
\end{figure}

The graphite plate carrying the detector was placed on the glovebox heater (setpoint 500~$^\circ$C). The Ge top-surface temperature was monitored with a surface probe and reached $\sim$280~$^\circ$C after $\sim$5~min. The sample was held at this surface temperature for 30~min under argon, then cooled rapidly on a large aluminum block to limit continued diffusion during the cooldown transient. Lithium diffusion is governed by temperature-dependent diffusivity and solubility limits, and rapid cool-down also helps control redistribution and precipitation behavior that can influence the final donor profile and inactive-layer properties~\cite{Fuller1953_LiDiff_GeSi,Pell1957_LiSolub_Ge,Morin1957_LiPrecip_Ge,Wenzl1971_LiPrecipitates_Ge,Crank1975_Diffusion}. After cooling to near room temperature, the diffused surface showed a uniform brick-red color.
The evolution of the surface appearance through deposition, diffusion, and cooldown is shown in Fig.~\ref{fig:LiPaintStages}.

We emphasize that this lithium-paint route is a non-standard step in modern HPGe fabrication; it is pursued here as a conformal alternative to one-step evaporation for future nonplanar electrodes. At present, KL01 represents the first fully operational hybrid planar detector fabricated at USD using this protocol; statistical yield and device-to-device variation will be established in a subsequent fabrication series (Sec.~\ref{sec:conclusion}).

\subsection{Post-diffusion cleaning and etching prior to sputtering}
\label{subsec:post_li_clean}
\begin{figure}
  \centering
  \includegraphics[width=0.8\columnwidth]{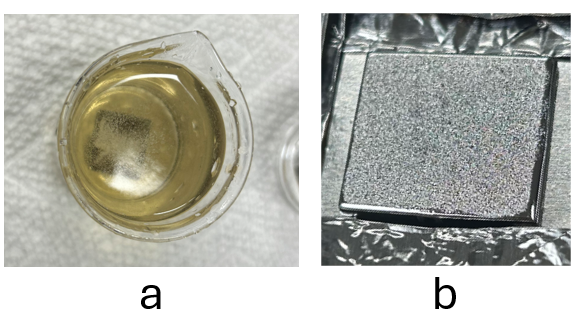}
  \caption{Lithium contact during and after the cleaning process. (a) Methanol bath immediately after immersing the detector with hydrogen bubble formation. (b) Lithium contact after cleaning.}
  \label{fig:LithiumSurface}
\end{figure}

Before sputter deposition, residual lithium and reaction products must be removed to prevent uncontrolled interfaces and high leakage. The cooled detector was removed from the glovebox and immediately immersed in methanol; vigorous bubbling was observed as lithium reacts to form lithium methoxide and hydrogen, providing a practical indication of active lithium removal. The surfaces were gently brushed to assist removal. After bubbling ceased, the detector was transferred to IPA, rinsed in deionized (DI) water, and dried with nitrogen. The methanol reaction and the post-cleaning surface condition are shown in Fig.~\ref{fig:LithiumSurface}.

\begin{figure}
  \centering
  \includegraphics[width=0.6\columnwidth]{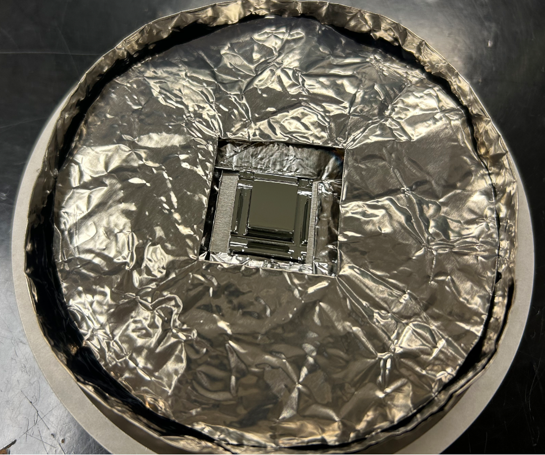}
  \caption{Detector mounted on the sputtering jig. The Ge crystal is positioned in the central opening, while aluminum foil covers surrounding jig surfaces to capture a-Ge overspray and minimize re-deposition.}
  \label{fig:SputterJig}
\end{figure}

A brief HNO$_3$:HF (4:1) etch ($\sim$30~s) was then used to remove the reacted surface layer and refresh the Ge surface prior to a-Ge deposition, following established practice for amorphous-contact HPGe fabrication~\cite{Wei2019_JINST_aGeDetectors,Amman2020_SegmentedAS}. The detector was rinsed thoroughly in DI water, dried with nitrogen, and mounted on a sputtering jig. Aluminum foil shielding was used to capture overspray and reduce re-deposition onto unintended surfaces, which is important for preventing parasitic conductive paths and maintaining sidewall passivation integrity~\cite{Amman2020_SegmentedAS}. The detector mounting geometry and aluminum-foil overspray shielding configuration are shown in Fig.~\ref{fig:SputterJig}.

\subsection{Sputtered a-Ge/Al contact and sidewall passivation}
\label{subsec:sputter_contacts}
Amorphous-Ge contacts are attractive because they can provide conformal passivation and effective blocking barriers at cryogenic temperature; their performance is sensitive to film microstructure, hydrogen incorporation, and interface condition, motivating controlled sputter parameters and careful vacuum practice~\cite{Wei2019_JINST_aGeDetectors,Bhattarai2020_EPJC_aGeConduction,Amman2020_SegmentedAS,Luke1992_aGeBipolar}. The AJA magnetron sputtering system used here employs a turbomolecular pump; therefore, the load-lock and main chamber were pumped overnight to reduce residual water. Deposition proceeded once the main-chamber base pressure reached $\sim 3\times10^{-7}$~Torr. The a-Ge target was pre-sputtered to clean the target surface prior to coating the detector~\cite{Wei2019_JINST_aGeDetectors,Amman2020_SegmentedAS}.

\begin{figure}
  \centering
  \includegraphics[width=0.6\columnwidth]{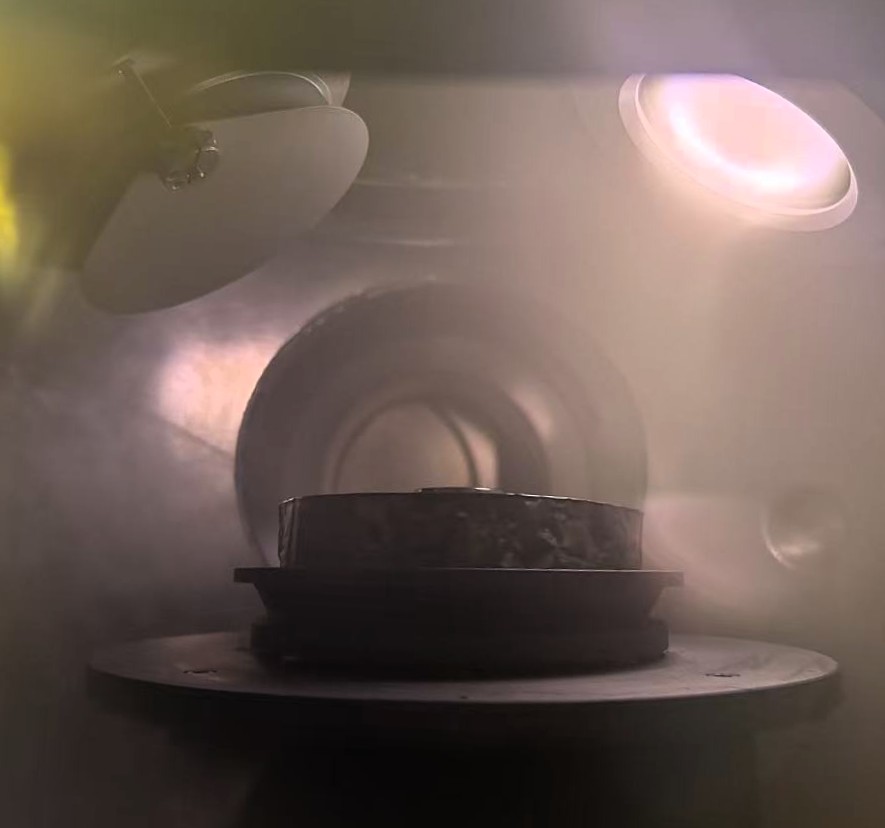}
  \caption{View inside the AJA sputtering chamber during amorphous-germanium deposition. The detector on the rotating stage is visible in the foreground and the a-Ge target/shutters are illuminated by the plasma glow.}
  \label{fig:aGeSputter}
\end{figure}

The a-Ge layer was deposited in a 7\% H$_2$ in Ar process gas with a total flow of 20~sccm, at 3~mTorr and 200~W RF power. The stage was rotated at 5~rpm to improve uniformity on the top surface and sidewalls. The 15~min deposition yielded an a-Ge thickness of $\sim$370~nm under these conditions. After deposition, the detector was cooled in vacuum to stabilize film temperature and reduce adsorption before metallization. A representative view of the chamber during a-Ge deposition is shown in Fig.~\ref{fig:aGeSputter}.

Without breaking vacuum, an aluminum film was deposited on top of the a-Ge layer to provide a low-resistance current-spreading layer and a practical interface to the readout chain, while the a-Ge/Ge interface supplies the injection-blocking behavior~\cite{Wei2020_EPJC_CBH,Luke1992_aGeBipolar}. The Al target was pre-sputtered (shutter closed) prior to use. For contact deposition, pure Ar at 3~mTorr and 20~sccm was used, with 400~W DC power applied for 6~min, yielding an Al thickness of $\sim$120~nm. Under this sequence, the a-Ge coats both the top face and sidewalls, producing sidewall passivation, while the Al capping layer is intended primarily for the top electrode.

After completing the top/side deposition, the chamber was vented and the detector was flipped and re-mounted so that the Li-diffused backside faced the target. An additional Al layer was deposited on the n$^+$ backside using the same Al sputtering conditions. This backside Al provides a low-resistance ohmic interface to the Li-diffused electrode, consistent with standard HPGe electrode practice where Li-diffused contacts are metallized to reduce series resistance and ensure stable high-voltage (HV) biasing~\cite{Knoll2010_RDM,Canberra2017_GeManual,Yin1969_Thesis_LiPaint}.

\subsection{Sidewall metal removal and finalization}
\label{subsec:sidewall_metal_removal}
To ensure that the sidewalls remain passivated (a-Ge only) and to prevent unintended wrap-around metal, the top and bottom electrodes were masked with vinyl tape and the detector was immersed in dilute HF (1\%) to remove Al from the side surfaces. Sidewall metal removal is important because metallic films on the sidewall can create parasitic electrodes, raise capacitance, and provide surface leakage pathways under high electric field~\cite{Wei2019_JINST_aGeDetectors,Amman2020_SegmentedAS}. The detector was rinsed in DI water and dried with nitrogen, and the etch was repeated as needed until no residual sidewall Al was visible. After tape removal, fabrication of KL01 was complete and the detector proceeded to electrical and spectroscopy characterization.

\section{Characterization of the Hybrid Detector and Mechanisms of the Li-Diffused Layer}
\label{sec:characterization}

This section summarizes the electrical and spectroscopic characterization of KL01 at 77~K and connects the observed performance to the underlying physics of (i) bulk depletion in a lightly doped p-type crystal, (ii) leakage-current mechanisms governed by carrier injection and surface conduction, and (iii) charge collection and inactive-layer effects associated with the Li-diffused n$^+$ contact. These considerations are central to HPGe detector operation and are discussed extensively in standard references and prior HPGe contact studies~\cite{Ma2017_ARI_InactiveLayer,Dai2022_arXiv_LiInactive,Knoll2010_RDM,Canberra2017_GeManual,Aguayo2013_NIMA_LiNplus,Haller1981_PhysicsUPGe}.

\subsection{Biasing, cryogenic operation, and leakage-current mechanisms}
\label{subsec:iv_setup}
The hybrid planar HPGe detector was mounted on the cold finger of the USD test cryostat, evacuated to high vacuum ($1.6\times10^{-6}$~mbar), and operated at liquid-nitrogen temperature (77~K). We define the bias polarity such that a positive high voltage is applied to the Li-diffused n$^+$ contact. The opposite a-Ge/Al electrode is held near ground and serves as the signal contact. This polarity corresponds to reverse-biasing an effective n$^+$--p junction, enlarging the depletion region in the lightly doped p-type bulk and reducing junction capacitance as bias increases~\cite{Knoll2010_RDM,Haller1981_PhysicsUPGe}.

For the I--V scan, the leakage current was measured with a Keithley picoammeter connected on the signal side. The picoammeter was connected to the a-Ge/Al electrode through a 1~G$\Omega$ series resistor; the high voltage was supplied to the n$^+$ contact through a low-noise filter. At 77~K, the dominant contributors to reverse current in HPGe devices are typically (i) thermionic/field-assisted carrier injection at imperfect contacts or barriers, (ii) generation-recombination in depleted regions, and (iii) surface leakage along sidewalls and edge regions; the latter can be strongly influenced by passivation quality and unintended wrap-around metallization~\cite{Wei2020_EPJC_CBH,Panth2020_EPJC_Cryo,Amman2020_SegmentedAS,Knoll2010_RDM}. The pA-level currents observed for KL01 (Sec.~\ref{subsec:cv_proxy}) are therefore a sensitive indicator that both the a-Ge passivation and the hybrid contact interfaces suppress injection and surface conduction under kV bias.

\subsection{Capacitance--voltage characteristics and depletion-voltage determination}
\label{subsec:cv_proxy}
For the C--V characterization we used a pulser-based capacitance estimate to track depletion behavior without a dedicated LCR bridge. A fixed-amplitude test pulser was injected and processed by the charge-sensitive preamplifier and a spectroscopy shaping amplifier; the shaped output was recorded with an oscilloscope. In practice, the observed pulser peak amplitude depends on the effective input capacitance of the detector-plus-front-end and therefore changes as the reverse bias expands the depletion region and reduces the junction capacitance. We use this response to construct an \emph{effective} capacitance-versus-bias curve (Fig.~9) that should be interpreted as an estimate (not a precision LCR measurement) with systematic uncertainty dominated by fixed stray capacitances and injection-path parasitics~\cite{Knoll2010_RDM,Mirion_HPGeLab}. The depletion voltage is identified by the transition to a clear plateau, where the capacitance is governed primarily by detector geometry plus approximately bias-independent contributions.

\begin{figure}[!htbp]
  \centering
  \includegraphics[width=0.8\columnwidth]{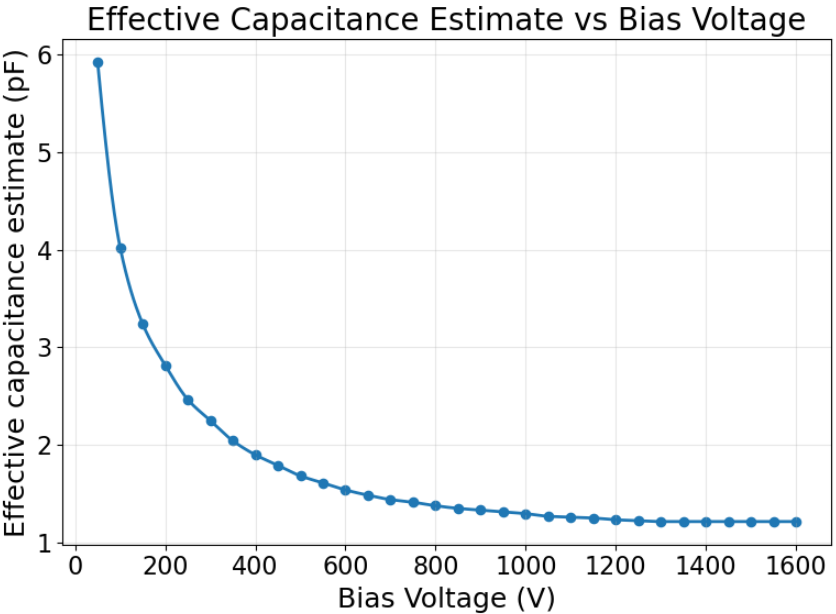}
  \caption{Effective capacitance estimate versus bias voltage for the hybrid planar HPGe detector at 77~K, obtained from the peak amplitude of the Gaussian-shaped test-pulser response measured at the shaping-amplifier output. The curve approaches a clear plateau above $\sim$1300~V, indicating full depletion; the effective capacitance scale should be interpreted as an estimate subject to fixed stray and injection-path parasitics.}
  \label{fig:cv}
\end{figure}

Figure~9 shows the effective capacitance estimate as a function of bias voltage at 77~K. The capacitance decreases rapidly at low bias as the depletion region expands and then approaches a clear plateau above $\sim$1300~V. We therefore identify $V_{\mathrm{dep}}\approx 1300$~V as the depletion voltage of the hybrid device and operate at 1600~V for spectroscopy to ensure stable, fully depleted operation. Across the scan range, the leakage current remains in the pA regime, increasing from $\sim$2~pA at 50~V to $\sim$15~pA at 1600~V, consistent with effective suppression of injection and surface leakage under high electric field~\cite{Wei2020_EPJC_CBH,Panth2020_EPJC_Cryo,Amman2020_SegmentedAS}.

The measured depletion behavior is also consistent with the expected electrostatics for a lightly doped planar HPGe diode. For a uniformly doped planar detector, the depletion voltage scales as $V_{\mathrm{dep}}\propto N_{\mathrm{eff}}d^2$~\cite{Knoll2010_RDM}, so kV-scale depletion is expected for $N_{\mathrm{eff}}\sim10^{10}$~cm$^{-3}$ and $d\sim$cm. In KL01, the presence of a Li-diffused inactive layer modifies the effective active thickness and can shift the apparent depletion plateau relative to the ideal bulk estimate; this effect is addressed quantitatively later via the Li profile and depleted-thickness estimate (Sec.~\ref{subsec:li_profile_analysis})~\cite{Ma2017_ARI_InactiveLayer,Dai2022_arXiv_LiInactive,Aguayo2013_NIMA_LiNplus}.

\subsection{Gamma spectroscopy, electronic noise, and experimental capacitance}
\label{subsec:spectroscopy_capacitance}
The spectroscopic performance was evaluated using a low-noise readout chain. Detector signals were read out with a charge-sensitive preamplifier based on the Lawrence Berkeley National Laboratory (LBNL) ``Square'' design topology. The input stage uses a BF862 JFET and a feedback capacitance $C_f=0.6$~pF, giving a nominal charge-to-voltage gain of $V_{\mathrm{out}}\approx Q/C_f$ for ideal charge integration~\cite{Knoll2010_RDM}. The preamplifier output was shaped by an ORTEC 671 amplifier with 1~$\mu$s shaping time, selected to optimize the signal-to-noise ratio among tested values. In HPGe systems, the optimum shaping time reflects the trade-off among series noise (capacitance-dependent), parallel noise (leakage-current dependent), and $1/f$ noise contributions~\cite{Knoll2010_RDM}.

Two $\gamma$-ray sources were used: $^{241}$Am and $^{137}$Cs. The amplifier gain was adjusted for dynamic range (coarse gain 1k for $^{241}$Am and 100 for $^{137}$Cs). A precision pulser was injected at the HV bias node (HV-side injection) to monitor electronic noise and to enable an independent capacitance estimate using a calibrated charge reference.

\begin{figure}[!htbp]
  \centering
  \includegraphics[width=0.8\columnwidth]{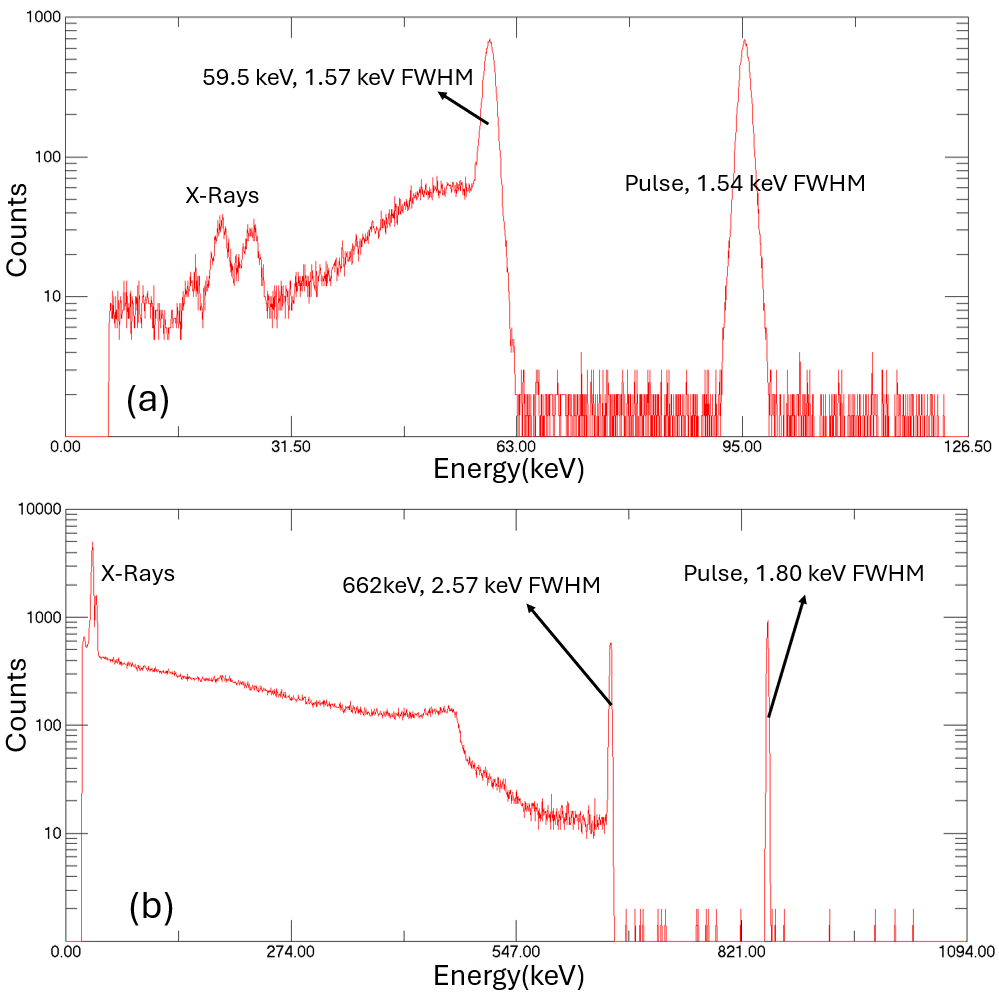}
  \caption{Energy spectra measured with the hybrid planar HPGe detector at 77~K, 1600~V bias, and 1~$\mu$s shaping time. 
  (a) $^{241}$Am spectrum showing the 59.5~keV peak and a pulser peak calibrated to 95.47~keV. 
  (b) $^{137}$Cs spectrum showing the 662~keV peak and a pulser peak calibrated to 852.94~keV.}
  \label{fig:spectra}
\end{figure}

Figure~\ref{fig:spectra}(a) shows the $^{241}$Am spectrum with the 59.5~keV line resolved with full width at half maximum (FWHM) $=1.57$~keV. The pulser peak (equivalent energy 95.47~keV) has FWHM $=1.54$~keV. Figure~\ref{fig:spectra}(b) shows the $^{137}$Cs spectrum with the 662~keV line at FWHM $=2.57$~keV and a pulser peak (852.94~keV) with FWHM $=1.80$~keV. The pulser widths quantify the electronic contribution under each gain configuration and therefore provide a direct handle on the noise floor of the readout chain in situ~\cite{Knoll2010_RDM}.

For Gaussian peak shapes, we convert FWHM to $\sigma$ by
\begin{equation}
  \sigma = \frac{\mathrm{FWHM}}{2.355}\,.
\end{equation}
A commonly used resolution parameterization in high-resolution $\gamma$ spectroscopy is the approximate $1/\sqrt{E}$ scaling,
\begin{equation}
  \frac{\sigma(E)}{E} = \frac{C[\%]}{\sqrt{E[\mathrm{MeV}]}}\,,
\end{equation}
which summarizes the dominant statistical broadening behavior and is widely used in HPGe calibration and benchmarking studies~\cite{Crespi2013_NIMA_AGATA15MeV,Giaz2012_EGAN_ResOverRootE}. Using the full-energy peaks, we obtain $C=0.27\%$ at 59.5~keV and $C=0.13\%$ at 662~keV.

To estimate the intrinsic detector contribution to the photopeak widths, we subtract the electronic component in quadrature:
\begin{equation}
  W_{\mathrm{int}} = \sqrt{W_{\mathrm{total}}^2 - W_{\mathrm{pulser}}^2}\,,
\end{equation}
yielding an intrinsic resolution of $\sim$0.30~keV at 59.5~keV and $\sim$1.83~keV at 662~keV. This indicates good charge collection in the depleted bulk and supports the conclusion that the hybrid contact scheme (Li n$^+$ plus a-Ge/Al) is operating stably at 77~K with low excess noise from leakage and surface effects~\cite{Wei2020_EPJC_CBH,Panth2020_EPJC_Cryo}.

\paragraph{Capacitance from pulser calibration.}
We also extract an experimental capacitance using the pulser calibration. For a pulser peak equivalent to $E=852.94$~keV, the collected charge is
\begin{equation}
  Q = \left(\frac{E}{\varepsilon}\right)e,
\end{equation}
where $\varepsilon=2.96$~eV is the mean energy required to form one electron--hole pair in Ge~\cite{HPGe2017_2p96eV}. Dividing by the measured preamplifier amplitude $V$ gives
\begin{equation}
  C_{\mathrm{exp}}=\frac{Q}{V}=1.215~\mathrm{pF}.
\end{equation}
A SolidStateDetectors.jl simulation for the realized geometry yields $C_{\mathrm{sim}}=1.667$~pF~\cite{Abt2021_JINST_SSD}. Both values are in the expected range for a compact planar HPGe device of $\sim$cm$^2$ area and $\sim$cm thickness. The $\sim$30\% difference is plausibly attributed to (i) the simulation's idealized boundary conditions and fully active electrode assumption, whereas KL01 includes a perimeter groove and handling wing that reduce the effective active field region, and (ii) systematic uncertainties in the pulser injection and readout chain (attenuation at the HV node, parasitic capacitance at the mount, and gain uncertainties)~\cite{Knoll2010_RDM,Abt2021_JINST_SSD}. Within a conservative $\sim$20\% systematic uncertainty budget, the experimental and simulated values are considered consistent.

\subsection{Benchmarking and implications for scaling Li-based hybrid contacts}
\label{subsec:benchmark_scaling}
To benchmark the quality of our Li-diffused n$^+$ contact, we compare to a recent report on indigenously developed planar HPGe detectors at the Bhabha Atomic Research Centre (BARC)~\cite{Pitale2023BARC}. In KL01, the depletion proxy shows a clear plateau above 1300~V and the leakage current remains in the pA regime up to 1600~V. In Ref.~\cite{Pitale2023BARC}, reverse-biased leakage currents below 100~pA at 700~V were used as a selection criterion for spectroscopic testing. The substantially lower leakage observed here, even at higher bias, indicates that the lithium-paint diffusion method can produce a low-noise n$^+$ contact with robust depletion behavior.

The measured energy resolutions, FWHM $=1.57$~keV at 59.5~keV and FWHM $=2.57$~keV at 662~keV, demonstrate stable operation and good bulk charge collection. Compared to the BARC report (e.g., $\sim$1.41~keV at 662~keV)~\cite{Pitale2023BARC}, the modestly larger FWHM in KL01 likely reflects a combination of electronics configuration and crystal/transport conditions (e.g., impurity level, trapping, or field nonuniformity), rather than a fundamental limitation of the hybrid contact concept. Importantly for scale-up, leakage-current suppression and stable high-voltage operation are the primary prerequisites for transferring the Li-paint approach to larger-area and nonplanar electrode topologies, where surface leakage and injection are typically the dominant failure modes~\cite{Ma2017_ARI_InactiveLayer,Dai2022_arXiv_LiInactive,Aguayo2013_NIMA_LiNplus}.

\subsection{Carrier concentration profiling of the Li-diffused layer via sequential Hall measurements}
\label{subsec:hall_profiling}
To determine the Li donor concentration profile in the n$^+$ layer, we employed sequential Hall effect measurements combined with controlled layer removal (lapping), a method suitable for profiling relatively thick doped layers and extracting active dopant concentration as a function of depth~\cite{Daubriac2018_DiffHall,Ghosh2025_LiGeContact}. A rectangular Ge coupon ($2.8$~mm $\times 1.5$~cm $\times 1.5$~cm) was cut from the same crystal and diffused under the same lithium-paint conditions as KL01, so that the extracted donor profile is representative of the n$^+$ contact of the hybrid device. At each lapping step, the sheet carrier density and Hall mobility were measured, with data taken at both room temperature and 77~K. The 77~K data are emphasized because reduced phonon scattering yields higher mobility and can improve the robustness of carrier-density extraction~\cite{Haller1981_PhysicsUPGe}.

This profiling approach is particularly relevant for Li-diffused HPGe because the final Li distribution depends not only on isothermal diffusion but also on cooling history, solubility limits, and precipitation/redistribution phenomena~\cite{Pell1957_LiSolub_Ge,Morin1957_LiPrecip_Ge,Wenzl1971_LiPrecipitates_Ge}. Consequently, an experimentally measured \emph{post-thermal-cycle} active donor profile provides a more realistic basis for interpreting dead-layer thickness and near-contact charge-collection behavior than an idealized constant-surface-concentration diffusion model alone~\cite{Ma2017_ARI_InactiveLayer,Dai2022_arXiv_LiInactive}.

\subsection{Lithium diffusion profile analysis and connection to dead-layer thickness}
\label{subsec:li_profile_analysis}
\begin{figure}[!htbp]
  \centering
  \includegraphics[width=0.8\textwidth]{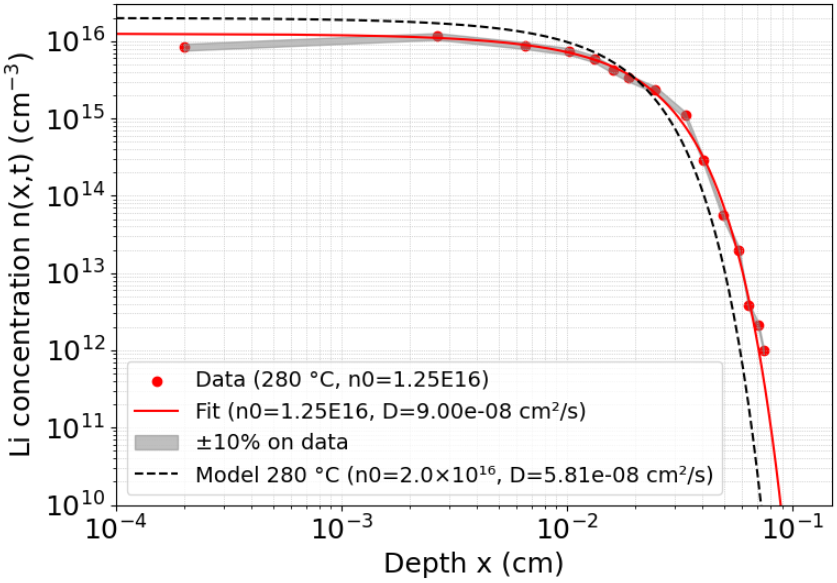}
  \caption{Lithium dopant concentration profile $n(x,t)$ in the n$^+$ contact as a function of depth $x$ after diffusion at 280~$^\circ$C for 30~min (both axes logarithmic). 
  Red points show Hall-effect data; the grey band indicates $\pm 10\%$. 
  The solid red curve is an error-function fit with effective parameters $n_0 \approx 1.25\times10^{16}$~cm$^{-3}$ and $D \approx 9.0\times10^{-8}$~cm$^2$/s. 
  The black dashed curve corresponds to the ideal diffusion profile using literature values $n_0 = 2.0\times10^{16}$~cm$^{-3}$ and $D = 5.81\times10^{-8}$~cm$^2$/s.}
  \label{fig:Li_profile}
\end{figure}

The Li-diffused n$^+$ layer can be approximated by diffusion theory governed by Fick's second law~\cite{Crank1975_Diffusion,Shewmon1989_Diffusion}. For a semi-infinite medium with constant surface concentration $n_0$, the solution is the complementary error function:
\begin{equation}
n(x,t)=n_0\,\mathrm{erfc}\!\left(\frac{x}{2\sqrt{Dt}}\right),
\end{equation}
where $D$ is the diffusion coefficient and $t$ is the diffusion time. Figure~\ref{fig:Li_profile} compares this idealized profile (using literature $n_0$ and $D$ values) to the Hall-profiled active donor distribution for the same recipe. The measured profile is broadly consistent with diffusion expectations, but modest deviations are visible in the tail region. Such deviations are consistent with the known sensitivity of Li distributions in Ge to precipitation and redistribution during cooling~\cite{Pell1957_LiSolub_Ge,Morin1957_LiPrecip_Ge,Wenzl1971_LiPrecipitates_Ge,Carter1960_LiPrecipitation}.

These profile details matter for detector performance because the Li-diffused region often constitutes an ``inactive'' layer: energy depositions near the n$^+$ contact can experience incomplete charge collection, distorted pulses, or degraded energies, depending on the local field and trapping in the Li-rich region~\cite{Ma2017_ARI_InactiveLayer,Dai2022_arXiv_LiInactive,Aguayo2013_NIMA_LiNplus}. This motivates quantifying the undepleted thickness on the Li side. Using the same planar depletion relation~\cite{Knoll2010_RDM,Mirion_HPGeLab}, and the bulk impurity concentration determined from a companion bipolar detector (fabricated from the same boule region), the depleted thickness at $V_{\mathrm{dep}}\approx1300$~V is estimated as $d\approx0.856$~cm. Comparing to the physical thickness (9.06~mm) implies $\sim$0.50~mm remains undepleted at the Li side. This estimate is consistent with the expectation that only the outer portion of the Li-diffused region behaves as a dead layer with respect to charge collection, while the p-type bulk beneath is fully depleted and active~\cite{Ma2017_ARI_InactiveLayer,Dai2022_arXiv_LiInactive,Aguayo2013_NIMA_LiNplus}.

\subsection{Barrier height of the Li n$^+$--p junction at 77~K}
\label{subsec:barrier_height}
To assess how sensitive the junction barrier is to the Li donor density, we estimate the built-in potential of an effective abrupt n$^+$--p junction,
\begin{equation}
\phi_B = \frac{k_B T}{q}\ln\!\left(\frac{N_DN_A}{n_i^2}\right),
\end{equation}
where $N_D$ is the donor density in the Li-diffused region, $N_A$ is the acceptor density in the p-type bulk, and $n_i$ is the intrinsic carrier concentration in Ge~\cite{Knoll2010_RDM,Haller1981_PhysicsUPGe}.
For KL01, we take $N_A \simeq 3.1\times 10^{10}$~cm$^{-3}$ from the depletion analysis of the bulk material. At 77~K, $n_i$ in Ge is extremely small~\cite{Haller1981_PhysicsUPGe}, so the logarithmic factor becomes very large and $\phi_B$ is only weakly dependent on order-of-magnitude variations in $N_D$ once $N_D \gtrsim 10^{14}$~cm$^{-3}$.

\begin{figure}[!htbp]
  \centering
  \includegraphics[width=0.8\textwidth]{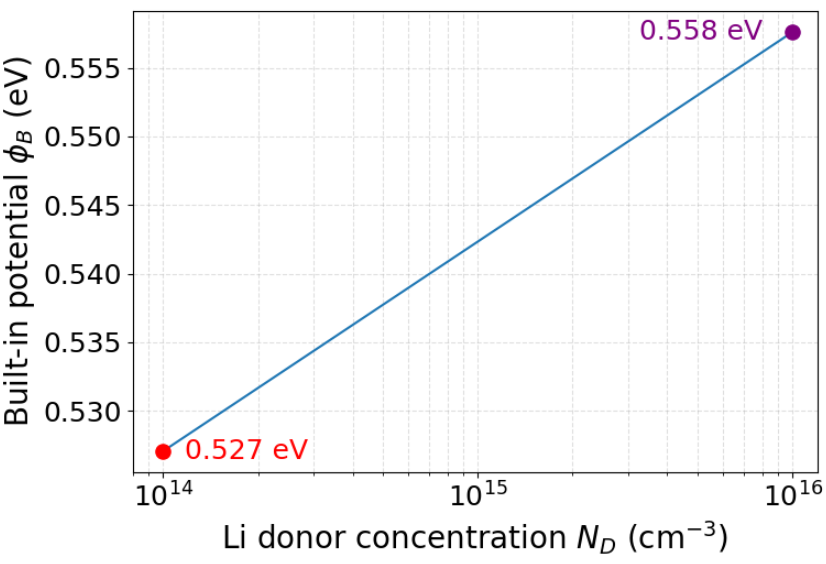}
  \caption{Calculated built-in potential $\phi_B$ of the effective Li n$^+$--p junction at 77~K as a function of Li donor concentration $N_D$ from $10^{14}$ to $10^{16}$~cm$^{-3}$, using $N_A=3.1\times10^{10}$~cm$^{-3}$ and an order-of-magnitude intrinsic concentration appropriate for Ge at 77~K~\cite{Haller1981_PhysicsUPGe}. The barrier varies only weakly across this range, consistent with robust rectifying behavior even when $N_D$ changes by orders of magnitude.}
  \label{fig:phiB_vs_ND_77K}
\end{figure}

Figure~\ref{fig:phiB_vs_ND_77K} shows $\phi_B$ at 77~K for $N_D$ spanning $10^{14}$--$10^{16}$~cm$^{-3}$. Over this two-decade range, $\phi_B$ increases only modestly (from $\approx 0.527$~eV at $10^{14}$~cm$^{-3}$ to $\approx 0.558$~eV at $10^{16}$~cm$^{-3}$ for the parameters above), demonstrating that a broad range of Li donor concentrations can still yield a robust rectifying barrier and low leakage at liquid-nitrogen temperature. In practice, variations in the \emph{shape} of the Li profile (and therefore the extent of the inactive/transition region) more strongly influence the effective dead-layer thickness and near-contact charge-collection behavior than the barrier height itself~\cite{Ma2017_ARI_InactiveLayer,Dai2022_arXiv_LiInactive,Aguayo2013_NIMA_LiNplus}.

For comparison, amorphous-Ge blocking contacts typically exhibit smaller effective charge barrier heights (order $\sim$0.3~eV) extracted from temperature-dependent leakage studies~\cite{Wei2020_EPJC_CBH,Panth2020_EPJC_Cryo}. This contrast highlights the complementary roles in the hybrid design: the Li n$^+$ electrode provides a robust high-voltage contact, while the a-Ge/Al signal side supplies strong injection blocking and stable passivation.

\section{Spectral Performance Comparison of Hybrid vs.\ Bipolar Detectors}
\label{sec:spectral_comparison}

\begin{figure}[!htbp]
\centering
\includegraphics[width=\linewidth]{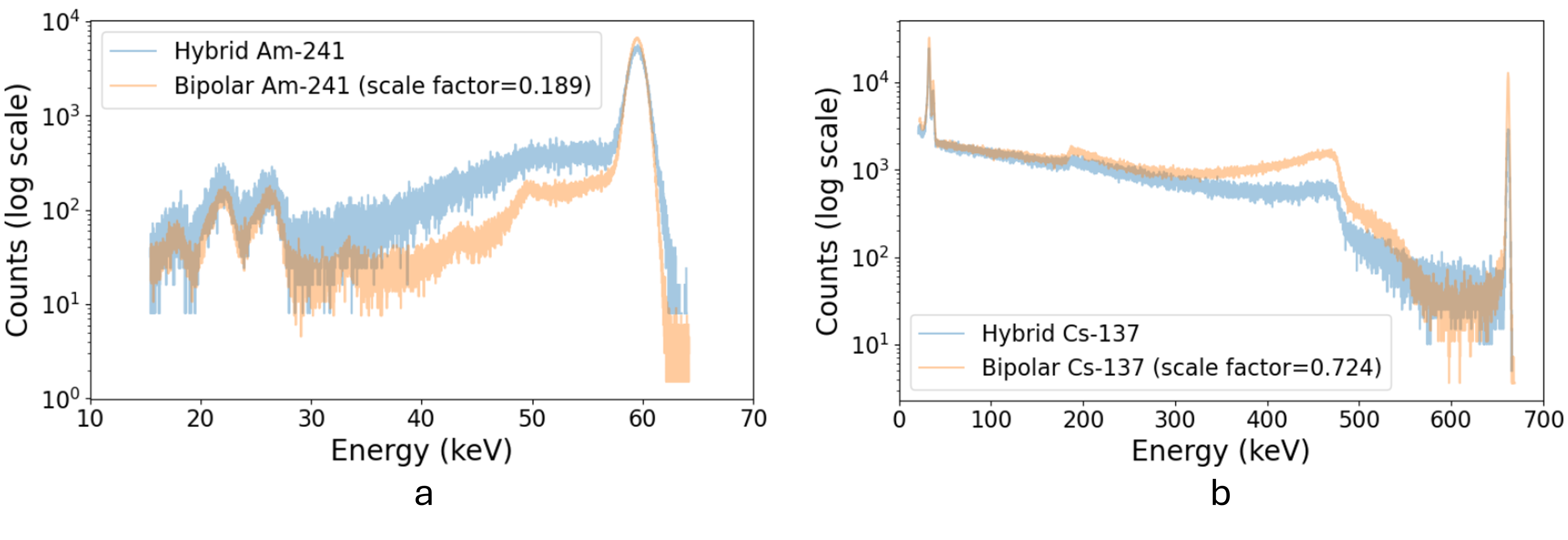}
\caption{Time- and volume-normalized $\gamma$-ray spectra of the hybrid Li/a-Ge planar detector (blue) and the bipolar a-Ge/a-Ge reference detector (orange), measured at 77~K with 1600~V bias and 1~$\mu$s shaping time for both detectors, and plotted on a logarithmic count scale. Panel (a) shows the $^{241}$Am spectrum in the low-energy region around the 59.5~keV line. Panel (b) shows the $^{137}$Cs spectrum over the 0--700~keV range, including the 662~keV full-energy peak and Compton continuum.}
\label{fig:hybrid_bipolar_spectra}
\end{figure}

To isolate the impact of contact architecture on spectroscopic response, we compared two planar detectors fabricated from USD-grown p-type HPGe: (i) the hybrid Li/a-Ge detector (KL01) and (ii) a bipolar a-Ge/a-Ge reference detector. Spectra from $^{241}$Am and $^{137}$Cs were recorded under identical operating conditions (77~K, 1600~V bias, identical preamplifier and shaping settings, and the same source-detector geometry). The spectra were normalized to acquisition time and to the \emph{total physical crystal volume} of each device. We intentionally normalized by the full geometric volume rather than the estimated active volume to treat the Li-diffused n$^+$ layer as a ``mass overhead'' intrinsic to Li-contact technologies. This choice yields a practical figure-of-merit for large-scale rare-event instrumentation: \emph{net full-energy response per unit mass of processed HPGe}, including any inactive material introduced by robust outer-contact implementations~\cite{Canberra2017_GeManual,Andreotti2014_ARI_DeadLayer,DAndrea2021_Universe_GeReview}. The resulting low-energy and medium-energy spectra are shown in Fig.~\ref{fig:hybrid_bipolar_spectra}, emphasizing differences in (i) full-energy peak widths, (ii) low-energy tailing, and (iii) continuum structure.

\paragraph{What differences are expected \emph{a priori}?}
The bipolar a-Ge/a-Ge device uses thin amorphous-Ge blocking contacts on both electrodes, so essentially the full crystal thickness can be depleted and active, while maintaining low leakage through bipolar blocking barriers and high-quality passivation~\cite{Panth2020_EPJC_Cryo,Wei2019_JINST_aGeDetectors,Amman2020_SegmentedAS,Luke1992_aGeBipolar}. In contrast, the hybrid device includes a Li-diffused n$^+$ electrode that produces a finite inactive layer (typically $\sim$0.5--1~mm in commercial HPGe, depending on diffusion recipe and subsequent processing)~\cite{Canberra2017_GeManual,Andreotti2014_ARI_DeadLayer}. In addition to the fully dead outer region, a ``transition'' zone can exist beneath the n$^+$ contact where the electric field is weak and/or the carrier transport is non-ideal; energy depositions there can exhibit incomplete charge collection and characteristic low-energy tailing~\cite{Ma2017_ARI_InactiveLayer,Dai2022_arXiv_LiInactive,Aguayo2013_NIMA_LiNplus,Gruszko2017_PhD_AlphaSurface}. Therefore, compared to a bipolar reference, a hybrid Li-contact detector is expected to show:
(i) reduced low-energy detection efficiency due to attenuation in the Li dead layer,
(ii) increased tailing from partial charge collection near the n$^+$ side, and
(iii) modestly degraded energy resolution when the degraded-energy population contributes under the photopeak~\cite{Ma2017_ARI_InactiveLayer,Knoll2010_RDM,Aguayo2013_NIMA_LiNplus}.

\paragraph{$^{241}$Am (59.5~keV) spectra: resolution, tailing, and efficiency.}
For the 59.5~keV line, the bipolar detector exhibits a sharper photopeak than the hybrid detector (Fig.~\ref{fig:hybrid_bipolar_spectra}a). The measured FWHM values of 1.57~keV (hybrid) and 1.46~keV (bipolar) correspond to $C \approx 0.27\%$ and $C \approx 0.25\%$ at 59.5~keV, respectively, using the $1/\sqrt{E}$ resolution scaling commonly adopted in HPGe spectroscopy~\cite{Crespi2013_NIMA_AGATA15MeV,Giaz2012_EGAN_ResOverRootE}.

Beyond FWHM, the dominant difference at 59.5~keV is the pronounced low-energy tail in the hybrid spectrum. We quantify this behavior using a compact tail-to-peak metric designed to be sensitive to events several $\sigma$ below the centroid (where partial collection contributes most strongly). With $\sigma=\mathrm{FWHM}/2.355$ evaluated for each spectrum, we define
\begin{align}
N_{\mathrm{peak}} &= \int_{\mu-2\sigma}^{\mu+2\sigma}\!\left[S(E)-B\right]\,dE, \\
N_{\mathrm{tail}} &= \int_{\mu-10\sigma}^{\mu-2\sigma}\!\left[S(E)-B\right]\,dE,
\end{align}
where $\mu$ is the photopeak centroid, $S(E)$ is the measured spectrum, and $B$ is a local continuum level estimated from the high-energy sideband
\begin{equation}
B=\left\langle S(E)\right\rangle_{E\in[\mu+3\sigma,\mu+5\sigma]}\,.
\end{equation}
The tail fraction is then \(f_{\mathrm{tail}} = N_{\mathrm{tail}}/N_{\mathrm{peak}}\).
For 59.5~keV we obtain $f_{\mathrm{tail}}=0.232$ (hybrid) and $f_{\mathrm{tail}}=0.117$ (bipolar), i.e.\ an approximately two-fold larger degraded-energy population in the hybrid detector.

Two physical mechanisms naturally explain this behavior. First, 59.5~keV photons are strongly attenuated by a sub-mm Li dead layer, so a larger fraction of interactions occur at, or near, the Li-contact side where incomplete charge collection is most likely. This effect reduces the full-energy peak yield \emph{per unit total crystal mass} and enhances the continuum at lower energies~\cite{Canberra2017_GeManual,Andreotti2014_ARI_DeadLayer}. Second, charge carriers produced in the n$^+$ transition region can experience recombination, trapping, and/or drift in weak-field regions, causing reduced collected charge and a continuum extending to lower energies~\cite{Ma2017_ARI_InactiveLayer,Dai2022_arXiv_LiInactive,Aguayo2013_NIMA_LiNplus}. This near-contact tailing mechanism is well documented for Li-diffused HPGe electrodes and is a key motivation for careful control of Li diffusion profiles and for modeling the ``inactive + transition'' layer as a function of processing history~\cite{Ma2017_ARI_InactiveLayer,Dai2022_arXiv_LiInactive,Aguayo2013_NIMA_LiNplus}.

\paragraph{$^{137}$Cs (662~keV) spectra: peak broadening and Compton-region differences.}
At 662~keV (Fig.~\ref{fig:hybrid_bipolar_spectra}b), both detectors show a clear full-energy peak and a Compton continuum. The bipolar detector again exhibits better resolution (FWHM $=2.05$~keV) than the hybrid detector (FWHM $=2.57$~keV), corresponding to $C \approx 0.11\%$ (bipolar) and $C \approx 0.13\%$ (hybrid). The hybrid detector also shows an enhanced population of counts just below the photopeak. Using the same tail-fraction definition, we find $f_{\mathrm{tail}}=0.0817$ (hybrid) and $f_{\mathrm{tail}}=0.0439$ (bipolar), again indicating roughly twice the relative tail population in the hybrid device.

At this higher energy, the role of simple attenuation in the dead layer is reduced, but charge-collection dynamics near the Li contact can still influence peak shape. Incomplete charge collection from interactions near the n$^+$ electrode can contribute to a low-energy tail and broaden the apparent photopeak if not fully separated by the fitting window~\cite{Ma2017_ARI_InactiveLayer,Dai2022_arXiv_LiInactive,Aguayo2013_NIMA_LiNplus}. In addition, the observed tailing can be amplified by electronics settings through \emph{ballistic deficit}: events with longer charge-collection times (e.g., those involving slow components near the Li contact or weak-field regions) can be underestimated in amplitude for a fixed shaping time, effectively shifting those events to lower apparent energies~\cite{Knoll2010_RDM,Aguayo2013_NIMA_LiNplus}. This provides a practical diagnostic: repeating the comparison at longer shaping times (and/or with digitized pulse-shape analysis) can help disentangle intrinsic charge-collection loss from shaping-time-dependent amplitude loss~\cite{Knoll2010_RDM,Aguayo2013_NIMA_LiNplus}.

\paragraph{Interpretation: contact-induced ``inactive + transition'' layer versus fully active bipolar surfaces.}
The spectral differences in Fig.~\ref{fig:hybrid_bipolar_spectra} are consistent with the known trade-off between thin, fully active amorphous-contact devices and hybrid devices employing a Li-diffused n$^+$ outer electrode. The bipolar detector represents a best-case planar benchmark because (i) both electrodes are thin and active, and (ii) a-Ge blocking contacts can provide low leakage with approximately symmetric barriers when properly prepared and passivated~\cite{Wei2020_EPJC_CBH,Panth2020_EPJC_Cryo,Wei2019_JINST_aGeDetectors,Luke1992_aGeBipolar}. In contrast, the hybrid detector contains a Li-diffused region that is partly inactive and partly transitional, enabling robust high-voltage operation but introducing charge-collection non-idealities for near-contact interactions~\cite{Ma2017_ARI_InactiveLayer,Dai2022_arXiv_LiInactive,Canberra2017_GeManual,Aguayo2013_NIMA_LiNplus,Andreotti2014_ARI_DeadLayer}.

This compromise is not merely a limitation: Li-diffused n$^+$ electrodes are central to large-mass HPGe technologies because they are mechanically robust, tolerant to handling, and able to sustain kV-scale biases over centimeter-scale depletion distances in kilogram-scale detectors, including low-background arrays~\cite{DAndrea2021_Universe_GeReview,Ackermann2013_EPJC_GERDA,Agostini2019_EPJC_BEGeChar}. Moreover, the thick n$^+$ layer provides powerful passive rejection of surface backgrounds: it fully stops $\alpha$ particles (range of only tens of $\mu$m in Ge) and strongly suppresses low-energy external radiation, which is advantageous for rare-event searches~\cite{Canberra2017_GeManual,DAndrea2021_Universe_GeReview,Gruszko2017_PhD_AlphaSurface,Barbeau2009_PhD}. The present hybrid-vs-bipolar comparison therefore provides a quantitative, process-relevant target for improvement: preserve the HV robustness and background advantages of Li n$^+$ while minimizing the thickness and charge-collection impact of the transition region through refined Li deposition/diffusion control and optimized surface finishing~\cite{Ma2017_ARI_InactiveLayer,Dai2022_arXiv_LiInactive,Aguayo2013_NIMA_LiNplus}.

\paragraph{Summary and near-term performance target.}
In summary, the bipolar a-Ge/a-Ge detector demonstrates the superior peak sharpness achievable when essentially the full crystal is active, while the hybrid Li/a-Ge device exhibits the expected signatures of a Li-contact ``inactive + transition'' region: reduced low-energy spectral purity and enhanced low-energy tailing at both 59.5~keV and 662~keV. The tail-fraction metric provides a compact benchmark for future iterations: reducing $f_{\mathrm{tail}}$ toward the bipolar baseline while maintaining pA-level leakage and stable depletion would indicate successful control of Li-layer mechanisms in the hybrid process flow~\cite{Ma2017_ARI_InactiveLayer,Dai2022_arXiv_LiInactive,Aguayo2013_NIMA_LiNplus}.

\section{Conclusion and Outlook}
\label{sec:conclusion}

We have fabricated and characterized the KL01 hybrid planar HPGe detector, representing the first fully operational planar device produced at USD that combines a lithium-diffused n$^+$ electrode with a sputtered amorphous-Ge/Al signal contact and a-Ge sidewall passivation. Operating at 77~K, KL01 reaches full depletion at $\sim$1.3~kV, maintains pA-level leakage current across the operating range (of order $10^{-11}$~A at the nominal bias), and achieves an energy resolution of 2.57~keV FWHM at 662~keV (corresponding to $\sim$0.39\% FWHM/$E$ and $C\simeq0.13\%$ in Eq.~(3)). These results place KL01 within the expected performance envelope of compact planar HPGe detectors and confirm that the USD-grown crystal quality and the implemented contact/passivation processes support stable, low-noise spectroscopy at liquid-nitrogen temperature~\cite{Panth2020_EPJC_Cryo,Wei2019_JINST_aGeDetectors,Wei2022_EPJC_PlanarPPC,Knoll2010_RDM,Haller1981_PhysicsUPGe}.

A central outcome of this work is the validation of a \emph{scalable process flow} rather than a single optimized detector. First, we demonstrate that the in-house lithium-paint deposition followed by controlled diffusion can form a functional, low-leakage n$^+$ electrode on a USD-grown crystal. Lithium-diffused contacts remain the most widely used outer-electrode technology for large HPGe detectors because they provide robust high-voltage operation and mechanically durable contacts, but they also introduce an inactive layer and a transition region where charge collection can be incomplete~\cite{Ma2017_ARI_InactiveLayer,Dai2022_arXiv_LiInactive,Canberra2017_GeManual,Aguayo2013_NIMA_LiNplus,Andreotti2014_ARI_DeadLayer}. KL01 provides a controlled test vehicle showing that the lithium-paint route can achieve the required electrical behavior (clear depletion plateau and pA leakage) while remaining compatible with subsequent wet processing and thin-film steps. Second, we confirm that the AJA sputtering recipe for a-Ge passivation and a-Ge/Al contacts produces stable, adherent films that suppress surface leakage and enable reliable readout from a thin signal electrode—capabilities that are essential when moving from small planar samples to larger surfaces and nontrivial topologies~\cite{Luke1992_aGeBipolar,Wei2020_EPJC_CBH,Amman2020_SegmentedAS}. Together, these accomplishments establish a repeatable fabrication and characterization baseline for scale-up in the USD detector pipeline.

The comparative study against a bipolar a-Ge/a-Ge reference detector further clarifies the remaining performance-improvement targets for hybrid Li-based devices. As expected from prior work on Li-diffused n$^+$ contacts, KL01 exhibits stronger low-energy tailing and a larger degraded-energy population than the fully active bipolar detector, consistent with partial charge collection in the n$^+$ transition region and shaping-time-dependent effects for slow components near the Li contact~\cite{Ma2017_ARI_InactiveLayer,Dai2022_arXiv_LiInactive,Knoll2010_RDM,Aguayo2013_NIMA_LiNplus}. This quantifies the engineering trade-off that motivates the present program: lithium contacts are indispensable for robust high-voltage operation and surface-background suppression in large HPGe modules, but the lithium inactive/transition layer must be controlled to preserve spectral purity, especially at low energies relevant to rare-event searches~\cite{Canberra2017_GeManual,DAndrea2021_Universe_GeReview,Gruszko2017_PhD_AlphaSurface}. The tail-fraction and resolution metrics reported here provide concrete benchmarks for subsequent process iterations.

\paragraph{Outlook toward scalable electrode geometries.}
The immediate next step is to translate the validated KL01 processes to advanced electrode designs that support kilogram-scale crystals. In particular, the ring-contact geometry proposed by Radford aims to preserve the low-capacitance advantages of point-contact readout while enabling substantially larger detector masses through field shaping with a ring-and-groove topology~\cite{Radford2024_RingContactTalk,Leone2022_Thesis_RingContact,Radford2018_PIRE_RingContact}. This direction is motivated by the broader trend in next-generation $0\nu\beta\beta$ programs: point-contact detectors and inverted-coaxial point-contact detectors enable powerful pulse-shape discrimination and have been deployed at the kilogram scale, while further increases in detector mass can reduce channel count and associated complexity in tonne-scale arrays~\cite{Agostini2021_EPJC_ICPC,Abgrall2021_LEGEND1000,DAndrea2021_Universe_GeReview,Ackermann2013_EPJC_GERDA,Agostini2019_EPJC_BEGeChar}. Demonstrating conformal lithium deposition and diffusion on nonplanar features is therefore a key risk-reduction milestone; the paint-and-diffuse approach explored here is specifically attractive because it can provide uniform coverage on vertical sidewalls and recessed grooves where one-step evaporation can be challenging.

\paragraph{Planned process and measurement improvements.}
Future work will focus on three parallel thrusts:
(i) \emph{Li-contact optimization and reproducibility:} fabricate additional KL-series hybrids to quantify yield and process window, and vary diffusion/cooling parameters to tune the Li profile and reduce the transition-region contribution to tailing~\cite{Ma2017_ARI_InactiveLayer,Dai2022_arXiv_LiInactive,Pell1957_LiSolub_Ge,Morin1957_LiPrecip_Ge,Wenzl1971_LiPrecipitates_Ge,Carter1960_LiPrecipitation};
(ii) \emph{surface and passivation refinement:} improve sidewall passivation uniformity and edge control (e.g., mitigating accidental metal wrap-around and suppressing surface conduction), building on established amorphous-contact fabrication protocols~\cite{Wei2020_EPJC_CBH,Wei2019_JINST_aGeDetectors,Amman2020_SegmentedAS};
and (iii) \emph{diagnostics that connect physics to processing:} repeat spectral comparisons at multiple shaping times and incorporate pulse-shape studies to separate intrinsic incomplete charge collection from ballistic deficit, thereby guiding geometry and process choices for future large-mass designs~\cite{Comellato2021_EPJC_Collective,Knoll2010_RDM,Aguayo2013_NIMA_LiNplus}.

In conclusion, KL01 establishes a practical and physics-informed fabrication baseline for hybrid-contact HPGe detectors at USD, bridging earlier small amorphous-contact prototypes and the larger-mass, field-shaped geometries needed by next-generation rare-event experiments. The demonstrated combination of pA-level leakage, stable depletion, and keV-scale energy resolution confirms the viability of the lithium-paint diffusion route when integrated with AJA-based a-Ge/Al contacts and sidewall passivation. With continued optimization and replication, these validated processes will be extended to ring-contact and related scalable architectures, advancing toward low-noise, large-mass HPGe modules relevant to LEGEND-1000 and other rare-event searches ~\cite{Abgrall2021_LEGEND1000,Radford2024_RingContactTalk,DAndrea2021_Universe_GeReview}.

\section*{Acknowledgements}
This work was supported in part by NSF OISE-1743790, NSF PHYS-2117774, NSF OIA-2427805,
NSF PHYS-2310027, and NSF OIA-2437416; by the U.S.\ Department of Energy under awards
DE-SC0024519 and DE-SC0004768; by the U.S.\ Air Force Office of Scientific Research under award
FA9550-23-1-0495; and by a research center supported by the State of South Dakota.
We gratefully acknowledge the USD Department of Chemistry for access to the glovebox facilities used in this work.
We thank Mark Amman for valuable technical advice on detector fabrication, and David C.~Radford and
Felix Hagemann for helpful guidance and discussions related to the simulations.
We acknowledge Lawrence Berkeley National Laboratory for providing the cryostat used for detector characterization in this work.

\bibliographystyle{unsrt}
\bibliography{refs.bib}

\end{document}